\documentclass[conference]{IEEEtran}
\IEEEoverridecommandlockouts
\usepackage{cite}
\usepackage{amsmath,amssymb,amsfonts}
\usepackage{algorithmic}
\usepackage{graphicx}
\usepackage{textcomp}
\usepackage{xcolor}
\def\BibTeX{{\rm B\kern-.05em{\sc i\kern-.025em b}\kern-.08em
    T\kern-.1667em\lower.7ex\hbox{E}\kern-.125emX}}
\begin{document}

\title{Generative AI for Analyzing Participatory Rural Appraisal Data: An Exploratory Case Study in Gender Research}

\author{
\IEEEauthorblockN{Srividya Sheshadri\IEEEauthorrefmark{1}, Unnikrishnan Radhakrishnan\IEEEauthorrefmark{2}, Aswathi Padmavilochanan\IEEEauthorrefmark{1}, \\ Christopher Coley\IEEEauthorrefmark{3}, Rao R. Bhavani\IEEEauthorrefmark{1}}
\IEEEauthorblockA{\IEEEauthorrefmark{1}Center for Women’s Empowerment and Gender Equality, Amrita Vishwa Vidyapeetham, Amritapuri, Kerala, India \\
Email: srividya.sheshadri@ammachilabs.org}
\IEEEauthorblockA{\IEEEauthorrefmark{2}Department of Business Development and Technology, Aarhus University, Herning, Denmark}
\IEEEauthorblockA{\IEEEauthorrefmark{3}Department of Social Policy and Data Sciences, Amrita Vishwa Vidyapeetham, Faridabad, India}
}

\maketitle

\begin{abstract}
This study explores the novel application of Generative Artificial Intelligence (GenAI) in analyzing unstructured visual data generated through Participatory Rural Appraisal (PRA), specifically focusing on women’s empowerment research in rural communities. Using the “Ideal Village” PRA activity as a case study, we evaluate three state-of-the-art Large Language Models (LLMs) --- GPT-4o, Claude 3.5 Sonnet, and Gemini 1.5 Pro --- in their ability to interpret hand-drawn artifacts containing multilingual content from various Indian states. Through comparative analysis, we assess the models’ performance across critical dimensions including visual interpretation, language translation, and data classification. Our findings reveal significant challenges in AI’s current capabilities to process such unstructured data, particularly in handling multilingual content, maintaining contextual accuracy, and avoiding hallucinations. While the models showed promise in basic visual interpretation, they struggled with nuanced cultural contexts and consistent classification of empowerment-related elements. This study contributes to both AI and gender research by highlighting the potential and limitations of AI in analyzing participatory research data, while emphasizing the need for human oversight and improved contextual understanding. Our findings suggest future directions for developing more inclusive AI models that can better serve community-based participatory research, particularly in gender studies and rural development contexts.
\end{abstract}


\section{Introduction}

The marginalization of women’s voices and experiences in many communities has long been a challenge in both research and policy-making for gender equality and sustainable development. Participatory Rural Appraisal (PRA) offers a transformative approach to addressing this exclusion, particularly in gender studies and community development \cite{b1}, \cite{b2}, \cite{b3}. By providing women with a platform to actively express their lived realities, PRA creates a space where their insights are centered and valued. Through the use of visual and narrative techniques, PRA enables women to articulate their unique social and economic vulnerabilities, which are often invisible or disregarded by traditional survey methods, and thus increases their representation in policies both locally and nationally. This participatory approach empowers women to contribute directly to decisions that affect their lives, ensuring their perspectives are heard and their challenges acknowledged \cite{b1}, \cite{b4}.  
While PRA is recognized for its ability to generate rich, context-specific data, the collection and analysis processes are resource-intensive \cite{b1}. The unstructured nature of the data—such as visual artifacts, narratives, and audio recordings—poses difficulties in scaling analysis efforts. PRA’s qualitative focus makes cross-site comparisons challenging, and variations in implementation and researcher skills can hinder data consistency. These limitations underscore the need for innovative tools that can streamline the analysis of PRA data while preserving its contextual richness. In this regard, Generative Artificial Intelligence (GenAI) offers a potential solution by automating the processing of large volumes of unstructured data. However, the unique nature of visual PRA data presents both opportunities and challenges for GenAI-based analysis.  
In recent years, GenAI has gained attention for its ability to analyze unstructured data, but its application remains primarily focused on mainstream sectors like business and finance \cite{b5}. Emerging areas like healthcare are beginning to leverage GenAI for analyzing unstructured, text-based information \cite{b6}. Despite existing frameworks for processing unstructured data \cite{b7}, they primarily address textual analysis, leaving visual data, particularly hand-drawn artifacts, underexplored. This is a critical limitation, given the growing need for GenAI tools capable of handling multi-modal forms of unstructured data, especially in applications of PRA within research domains such as empowerment and gender studies, where visual representation is often central to community-driven processes.   
The lack of GenAI tools to process visual hand-drawn data not only highlights a gap in current research but also poses a challenge for extending GenAI’s utility in grassroots-level participatory methodologies. It places a spotlight on the systemic gap between the promise of emerging technologies and the applicability of such for the populations that may benefit the most from them. Especially in the context of sustainable development for below-poverty-line communities, notable for a general lack of technology access and resources, the application of more advanced tools faces significant hurdles.  
To our knowledge, the application of AI to analyze unstructured PRA-based data, particularly in the context of empowerment, gender studies, and community development, remains unexplored. This study aims to address this gap by pioneering the analysis of unstructured PRA-based visual data through AI, advancing both AI applications and gender research. Specifically, it investigates the question: How can AI be leveraged to analyze unstructured, visual hand-drawn PRA data within the domain of gender research?  
To explore this, we adopt a case study approach, examining unstructured visual data from the \textit{Ideal Village} activity, a PRA tool used to investigate empowerment. By exploring AI’s potential to analyze such data, this study proposes a new direction for AI research with a focus on gender and development contexts. The findings have the potential to advance AI’s technical capabilities while promoting more equitable methodologies in gender and rural development research, supporting deeper insights into community-driven processes. 

\section{PRA: Ideal Village}

The \textit{Ideal Village} (\textit{IV}) activity is one of several PRA tools used to assess empowerment, following the framework set by the Advancing Women’s Empowerment through Systems-Oriented Model Expansion (AWESOME)\cite{b8}. The AWESOME framework addresses key challenges in women’s empowerment research, including the neglect of contextual complexities, lack of effective measurement tools, and unsustainable implementation efforts. It adopts a systematic, holistic approach to conceptualize empowerment as a dynamic interaction between various life dimensions (e.g., education, livelihood, health) and domains (e.g., access, opportunities, awareness, mental space) across individual, household, and community contexts with a focus on addressing gender-specific barriers and dynamics.   
Grounded in principles of psychological empowerment and transformative learning, the \textit{IV} activity fosters critical thinking, decision-making, and social problem-solving skills \cite{b9}. Serving as both a measurement tool and a project-based learning activity, it assesses empowerment at the community level while simultaneously activating the empowerment process. The \textit{IV} activity was leveraged to elucidate key concepts of the AWESOME framework. Through the \textit{IV} activity, we can gain insights into the following elements of the AWESOME Framework:

\begin{itemize}
    \item \textbf{Identifying key dimensions of life:} The activity helps participants identify the most pressing social-cultural, environmental, health, and economic issues affecting their community. This aligns with the dimensions of life in the AWESOME framework.
    
    \item \textbf{Assessing and activating empowerment domains:} The activity provides insights into the participants’ awareness, access, opportunities, and mental space (i.e., perceptions of self-belief) related to achieving their ideal village. The activity also activates the participants’ empowerment process via the mental space, by engaging them in critical reflection and problem solving to envision a different future.
    
    \item \textbf{Examining community-level factors:} The activity reveals the influence of broader community-level factors, such as social norms, cultural practices, and institutional structures, on individual and household-level empowerment.
    
    \item \textbf{Uncovering vulnerabilities:} The gap between the ideal and current village can highlight vulnerabilities in various dimensions and domains of empowerment, providing insights into the challenges faced by the community.
\end{itemize}

To ensure that the \textit{IV} activity yields these intended insights, it is crucial that it is implemented with fidelity and adherence to the prescribed protocol.  
Participants work in small groups of 10 or less to envision their ideal village and collectively illustrate this vision on chart paper. This collaborative activity engages their intrapersonal and interactional capacities, core elements of the Psychological Empowerment model \cite{b10}. They then critically reflect on the differences between their current and ideal villages, highlighting gaps and vulnerabilities in their community. This reflection fosters transformative learning, as participants confront these differences through open dialogue, gaining self-awareness and a deeper understanding of their shared and individual visions \cite{b11}.  
Next, participants explore the Circle of Control (CC), a tool that helps them categorize issues based on their ability to influence change. They reflect on whether certain issues are within their control (Circle of Control), can be influenced with help (Circle of Influence), or seem beyond their control (Circle of Concern). This exercise encourages them to reassess their power to affect change, particularly when working together as a community. The activity also helps the community to better articulate the support they need from local government and other decision-making stakeholders, and prepares them for seeking help through the appropriate channels. Through dialogue and consensus, participants prioritize one issue to address collectively, developing an action plan. This final phase goes beyond planning, as participants take concrete steps toward realizing their ideal village.

\section{AI for PRA}

Developing Generative AI models that incorporate diverse datasets and address intersectionality is recognized as critically important for mitigating bias and enhancing accuracy across varied demographic groups. The importance of inclusivity in AI is widely recognized, with studies highlighting that a lack of gender diversity in tech teams can lead to AI models that perpetuate sexist or prejudiced biases \cite{b12}. This is a reflection of broader societal inequalities, as noted by Borenstein and Howard \cite{b13} who argue that AI, if not developed with care, can exacerbate existing gender disparities, particularly in the Global South. While frameworks like those proposed by Halevy et al. \cite{b14} aim to detect and mitigate toxic language in AI, the challenge of addressing bias in non-text-based unstructured data, such as visual or hand-drawn artifacts, remains largely unexplored. Mahadevkar et al. \cite{b7} provide a framework for processing unstructured data but focus primarily on text-based analysis, highlighting a significant gap in the literature regarding the multi-modality of unstructured data. Their work suggests that existing methodologies for analyzing mixed data are insufficient for visual contexts, an area ripe for further investigation, particularly in fields like implementation science, where AI can bridge the evidence-to-practice gap \cite{b15}.  
   
\subsection{Generative AI}

Large Language Models (LLMs) are a subset of deep learning-based Generative AI models designed to process and generate human-like text \cite{b17}. LLMs are trained on massive datasets of natural language, allowing them to understand context, produce coherent responses, and generate insights from textual prompts. Their applications range from generating summaries of large amounts of text to simulating personas while having text conversations with humans. LLMs can also handle multilingual prompts, as was required for this study, which involved images containing text in English, Hindi, and Malayalam \cite{b18}. A subset of LLMs are also Multimodal Language Models (MLLMs), such as those GPT-4 and Gemini Pro, which extend the capabilities of LLMs by allowing models to analyze and interpret visual data alongside text \cite{b17}. In the context of our research, MLLMs are crucial for examining the hand-drawn artifacts from PRA activities. These models are capable of identifying visual elements (such as environmental features, houses, and other infrastructure) and linking them to textual descriptions or categories (e.g., social, economic, or health-related). Instructions to these models are called prompts. Prompt engineering refers to the iterative process of designing input prompts that guide LLMs toward producing desired outputs. Since Generative AI outputs are directly influenced by the structure and specificity of the prompts, multiple versions were tested in our study to improve the accuracy of the analysis. One of the main issues found in Generative AI models is that of “hallucinations”, when LLMs generate information not grounded in the input data \cite{b17}. This phenomenon can lead to erroneous interpretations, particularly in visual analysis, where the model may misidentify or over-interpret elements in an image. Managing hallucinations was a critical aspect of our analysis, as the accuracy of the AI interpretation had to be validated against expert human interpretation. 

\section{Methodology}

Our study employed three state-of-the-art Large Language Models (LLMs) to analyze PRA-generated visual artifacts: OpenAI’s GPT-4o (version: gpt-4o-2024-08-06), Anthropic’s Claude 3.5 Sonnet (version: claude-3-5-sonnet-20240620), and Google's Gemini 1.5 Pro (version: gemini-1.5-pro-exp-0827). These models were selected based on their superior performance on relevant benchmarks including MMLU and the \textit{lmsys} arena leaderboard. To optimize model performance while maintaining interpretive capabilities, we implemented standardized parameters across all three models: a temperature setting of 1.0 and an output token limit of 4095 (approximately 2500-3000 English words). The temperature parameter, which controls model creativity, was set at 1.0 following preliminary testing that showed minimal variation in hallucination rates between temperature settings of 0 and 1.  
\subsection{Prompt Development}

The development of prompts followed an iterative refinement process aimed at maximizing accuracy and consistency of model outputs. Our system prompts were standardized across all three models to ensure comparative validity, incorporating frameworks for element identification, dimensional categorization, and uncertainty reporting. The prompt structure evolved through multiple iterations, with each revision addressing specific challenges encountered during testing. The final implementation inspired by prompt engineering techniques such as chain-of-thought reasoning and the use of XML tags for structuring output \cite{b17} included comprehensive instructions for element identification, uncertainty handling protocols, and standardized output formatting requirements.  

\subsection{Datasets}

Our dataset comprises photos of drawings created as a part of the PRA, IV Activity—depicting ideal villages and circles of control. These drawings were made on chart paper by groups of women from rural communities across multiple Indian states, including Kerala, Jharkhand, Odisha, Uttarakhand, and Uttar Pradesh. Despite careful documentation protocols followed by PRA facilitators, some technical limitations emerged in the digital preservation of these artifacts. The technical specifications of these images varied considerably, with resolutions ranging from 960x1280 to 4032x3024 pixels, mixed format types including JPEG and PNG, and variable image quality due to variation in camera quality as well as compression from digital sharing platforms. The diversity in format, resolution, and language resulted in significant variation in content characteristics.

The content variability presented several analytical challenges. The linguistic diversity spanned English, Hindi, and Malayalam, sometimes with mixed-script and mixed-language annotations within single images. Drawing styles ranged from basic sketches to detailed illustrations, while handwriting quality varied from clearly legible to challenging scripts. The density of annotations also showed significant variation, from sparse labels to complex, multi-layered annotations.

\section{Results}

\begin{figure}[htbp]
\centerline{\includegraphics[width=\linewidth]{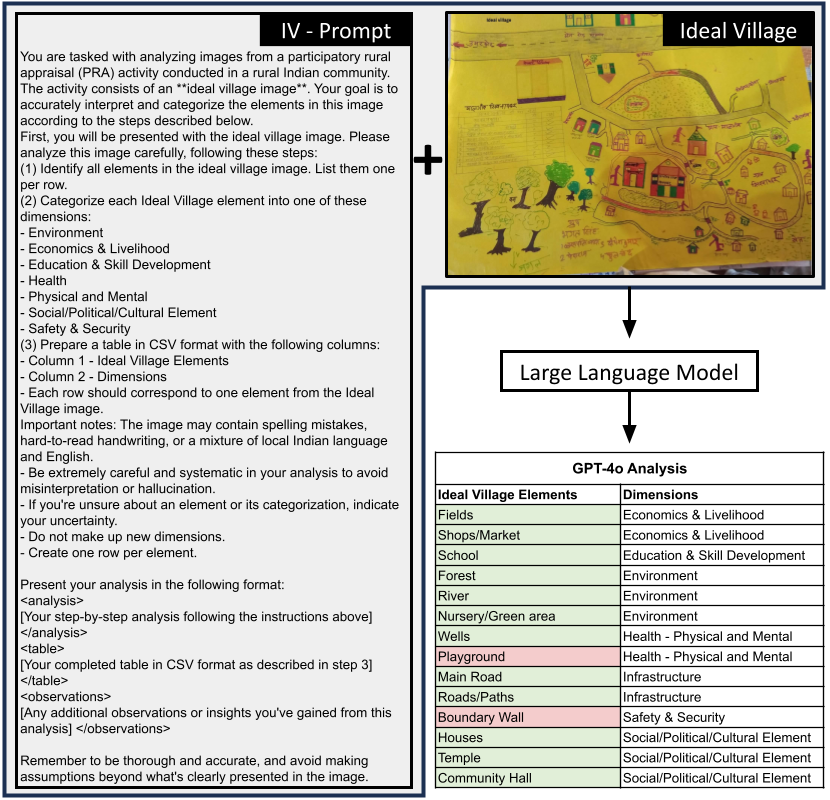}}
\caption{Sample output demonstrating GPT-4o's analysis of an Ideal Village participatory drawing from a rural Indian community. The analysis was generated using a structured prompt designed to identify and categorize village elements. The output table shows element classification across different dimensions, with green-highlighted rows indicating accurate interpretations and red-highlighted rows showing misclassifications or potential hallucinations by the model.}
\label{fig:iv_fig}
\end{figure}

\begin{figure}[htbp]
\centerline{\includegraphics[width=\linewidth]{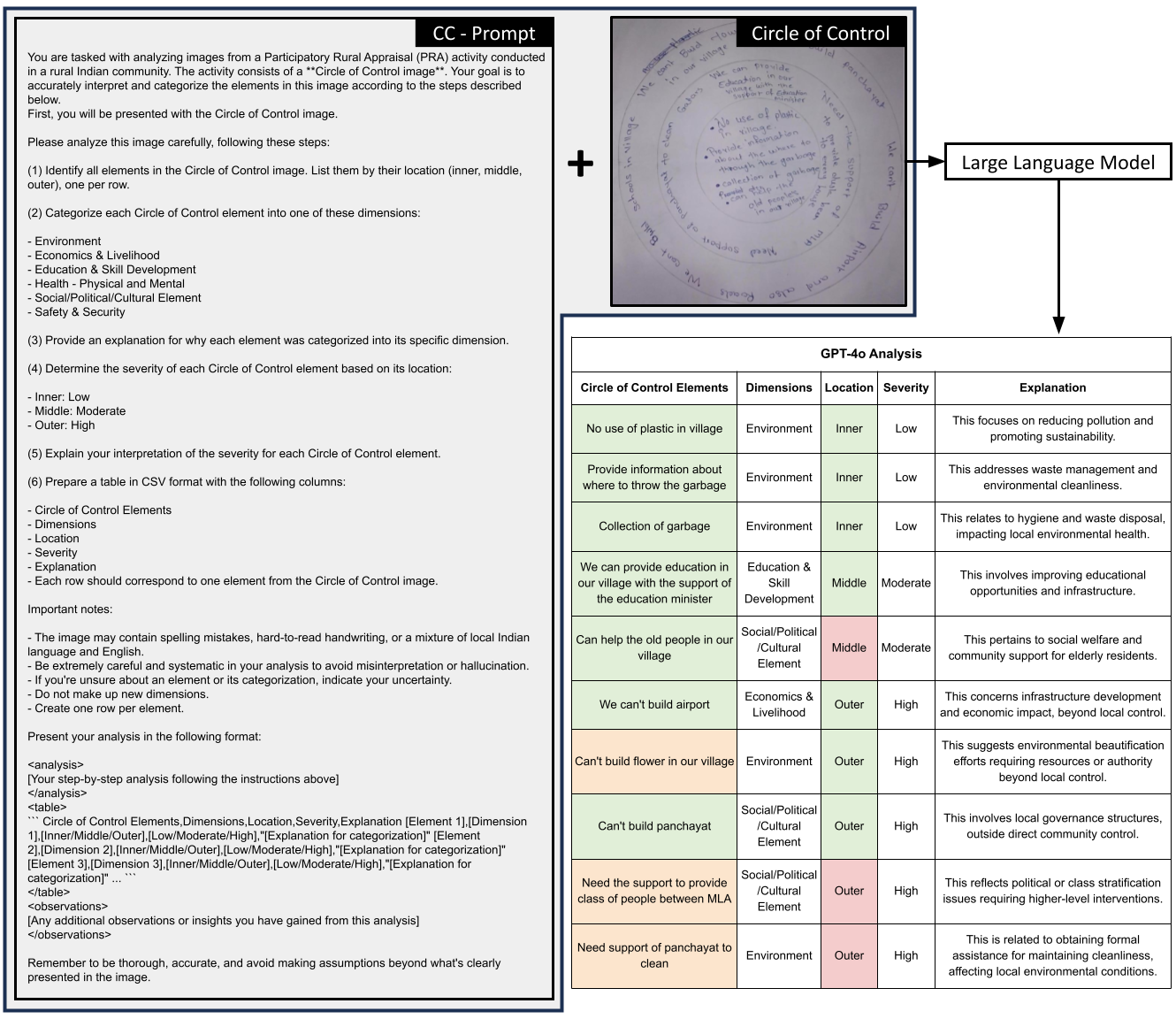}}
\caption{Sample output illustrating GPT-4o’s analysis of a Circle of Control participatory drawing from a rural Indian community. The model identified elements and their locations (inner, middle, or outer circles) while categorizing them across different dimensions. In the output table, green-highlighted rows indicate accurate interpretations across all columns, orange-highlighted rows show partially correct classifications, and red-highlighted rows indicate misclassifications.}
\label{fig:cc_fig}
\end{figure}

The prompts were designed to enable systematic analysis of both IV and CC drawings through a structured analytical framework. For IV analysis, the prompt required element identification and categorization across the AWESOME Framework dimensions including Environment, Economics \& Livelihood, Education \& Skill Development, Health, Physical and Mental, Social/Political/Cultural Element, and Safety \& Security. The CC prompt extended this framework by incorporating spatial analysis (inner, middle, outer circles) and severity assessment based on element location. Both prompts mandated specific output formats: IV utilized a two-column structure capturing elements and their dimensions, while CC employed a five-column format including location, severity, and explanatory documentation. The prompts incorporated protocols for handling multilingual content and maintaining analytical consistency across diverse PRA artifacts.   
The prompts were executed through a custom Python script designed to systematically process our dataset comprising 10 IV and 10 CC drawings through the three selected LLMs. IV images were analyzed using the \textit{IV - prompt} (Figure \ref{fig:iv_fig}), while CC images were processed using the \textit{CC - prompt} (Figure \ref{fig:cc_fig}). Model responses were automatically parsed and structured into CSV format, maintaining consistent categorization of elements across predefined dimensions. Samples of the prompts and responses are shown in Figures \ref{fig:iv_fig} and \ref{fig:cc_fig}. The prompt responses included step-by-step instructions and the PRA visual artifacts — the IV drawing and CC drawing respectively. The IV-Prompt output table (Figure \ref{fig:iv_fig}) contains elements detected in the drawing (visual, text, and/or numerical) along with their associated dimension classification. The CC-Prompt output table (Figure \ref{fig:cc_fig}) similarly includes the detected elements, their associated dimension classifications, the location and severity of these elements within the circles of control, and an explanation for each classification. An assessment was then conducted to evaluate the accuracy of element identification by comparing the output to the original visual sample.   
Due to the still evolving capabilities of cutting edge LLMs in understanding Indic scripts and complex spatial analysis, we see some interesting hallucinations exhibited by the models. An illustrative example is found in Figure \ref{fig:cc_fig} for instance, where GPT-4o analyses the input CC image. In a case of partially incorrect analysis, the CC item \textit{``Can't build flower in our village''} is identified, however the ‘flower’ element likely refers to a ‘tower’ due to potential handwriting ambiguity. Similarly, in the identified item \textit{``Need support of panchayat to clean''}, the model missed the term 'gatars' (likely meaning \textit{``gutters''}) at the end of the phrase. In one case a CC element is accurately identified: \textit{``Can help the old people in our village''}, but the model incorrectly identifies the location of it as 'middle' instead of 'inner' circle.   

In the context of PRA and gender studies, Table \ref{tab:performance} provides a comparative evaluation of the three LLMs explored in this study — GPT-4o, Claude 3.5 Sonnet, and Gemini 1.5 Pro — on their ability to handle unstructured visual, and multilingual data, that is typical of PRA research. The LLMs were evaluated on seven critical themes that directly affect GenAI’s efficacy in accurately capturing the community context, and preserving nuanced perspectives. The table rates each model on a high, medium, or low scale across themes like visual data interpretation, language and translation challenges, misclassification of data, and handling ambiguity—all of which are essential for analyzing PRA data accurately.   
All three models encountered significant challenges in interpreting visual elements, particularly when images were unclear. Clearer segmentation and higher resolution visuals generally improved performance across all models, whereas unclear or handwritten content frequently resulted in misclassifications or an over-reliance on background prompts, rather than accurate interpretation of the visual data. Visual interpretation difficulties were a pervasive issue. Even when the LLMs performed well with moderately clear and simplistic visuals, they still misinterpreted key elements. This highlights a critical limitation: inaccurate visual interpretation by GenAI can lead to the loss of essential context, particularly regarding gender-specific vulnerabilities.  
Language and translation challenges are also pertinent to PRA, which collects data in local languages and takes significant resources to accurately translate into English. In the case of this study, the models struggled with translating Hindi and Malayalam accurately, especially when interpreting unclear handwriting or dialects. In a gender studies context, translation accuracy is vital for preserving participant’s narratives, as mistranslation can obscure or distort the socio-economic realities being expressed.  
Misclassification of data was another notable theme where GPT and Claude performed poorly, both receiving high difficulty ratings. Misclassifications—such as associating schools with economic livelihoods or lumping unrelated elements together—can lead to skewed data interpretation, misrepresenting the priorities of women in rural areas. In contrast, Gemini, while still facing challenges, was better at capturing certain unique elements, like \textit{``respect for girls''}, which went unnoticed by the other models. The importance of accurate classification is especially relevant in gender studies, where the nuances of social roles, access to resources, and community dynamics are often intricately linked to women’s empowerment.  
Finally, across all models, hallucinations and assumptions emerged as a recurring pattern. Models introduced new elements or dimensions, indicating an inability to follow the prompt instructions. Additionally, the occurrence of hallucinations increased when the provided imagery was insufficient or of poor quality. In one instance for example, the element \textit{``dowry''} was inaccurately identified in one sample, accompanied by a fabricated explanation and an assessment of its severity by the community. In the context of PRA, such hallucinations can severely impact the understanding of a community.  
To conclude, while GenAI presents promising opportunities for automating and scaling the analysis of unstructured visual PRA data, the challenges identified in this matrix—particularly around translation, visual interpretation, and misclassification—underscore the need for careful human oversight. PRA’s focus on capturing the lived realities of women in rural settings necessitates an approach where GenAI aids rather than undermines or discredits the richness of the data.

\begin{table}[htbp]
\caption{Comparative performance of GPT, Claude, and Gemini on key themes relevant to the analysis of Participatory Rural Appraisal (PRA) data and gender studies. The table evaluates the models based on their ability to handle challenges such as language translation, visual data interpretation, misclassification, dimensional flexibility, and handling ambiguity. Ratings of High, Medium, or Low indicate the degree of difficulty or success each model experienced in processing qualitative, multilingual, and visual data.}
\begin{center}
\resizebox{\columnwidth}{!}{%
\begin{tabular}{|c|c|c|c|}
\hline
\textbf{Theme} & \textbf{GPT-4o} & \textbf{Claude 3.5 Sonnet} & \textbf{Gemini 1.5 Pro} \\
\hline
Language and Translation Challenges & Medium & High & Medium \\
\hline
Visual Data Interpretation Challenges & High & Medium & Medium \\
\hline
Misclassification of Data & High & High & Medium \\
\hline
Inconsistency Across Models & High & Medium & Medium \\
\hline
Handling of Ambiguity and Uncertainty & Medium & Medium & Medium \\
\hline
Human Oversight and Dataset Quality & High & High & High \\
\hline
Hallucinations and Assumptions & Medium & High & High \\
\hline
\end{tabular}%
}
\label{tab:performance}
\end{center}
\end{table}

Finally, in analyzing the Ideal Village PRA through GenAI models with the aim of extracting insights aligned with the AWESOME framework, several key observations emerged:

\begin{itemize}
    \item \textbf{Identifying Key Dimensions of Life:} The models demonstrated partial success in identifying essential empowerment dimensions. However, issues such as hallucinations and misclassifications were noted, impacting the reliability of insights.  
    \item \textbf{Assessing and Activating Empowerment Domains:} In this area, the models exhibited limited competency, likely due to inadequate data quality and quantity, which restricted the validity of outputs related to empowerment.  
    \item \textbf{Examining Community-Level Factors:} With the available data, the models generated preliminary assessments of community-level factors, though the depth and accuracy of these assessments were variable.  
    \item \textbf{Uncovering Vulnerabilities:} The models similarly produced some assessments of vulnerabilities within the community, though these outputs were constrained by data limitations.
\end{itemize}

\section{Discussion \& Limitations}

This study explored current GenAI capabilities in analyzing unstructured visual data within PRA and gender studies research. While the models showed competence in following prompts, they struggled with multilingual and qualitative data interpretation crucial to PRA analysis.

Our findings highlight significant dataset limitations, particularly the absence of contextual information through field notes. This reliance on visual imagery with minimal textual input hindered meaningful data interpretation, suggesting the need for more comprehensive datasets that include diverse contextual elements. These limitations also reflect broader challenges in conducting consistent PRA activities, which require expert facilitation.

The models exhibited frequent errors and hallucinations due to language barriers and visual misinterpretations, emphasizing the need for improved multilingual capabilities \cite{b18} and manual validation. Data quality posed additional challenges, with low-resolution images and insufficient community context limiting the models' effectiveness. Furthermore, our current prompt engineering approach would benefit from a more systematic evaluation framework incorporating parameters such as truthfulness, instruction following, and contextual relevance. Such refinements could better identify specific challenges in processing PRA data and guide targeted improvements in prompt engineering for gender research applications.

\section{Conclusion}

This study explored the application of GenAI in analyzing unstructured visual data, specifically hand-drawn artifacts. The findings indicate that AI is still in its early stages of visual data processing, revealing an urgent need for more inclusive models capable of handling diverse data types while addressing the ethical concerns associated with their deployment. By pioneering the analysis of PRA-based unstructured visual data, this research expands the boundaries of GenAI’s capacity to process and interpret complex, culturally rich representations that extend beyond conventional data types like text.  
As the analysis of unstructured data continues to evolve, the risk of exacerbating biases—particularly when transitioning from textual to visual artifacts—becomes increasingly significant. The implications of this research could be far-reaching, enhancing not only the technical capabilities of GenAI but also contributing to more equitable methodologies in gender research and rural development.  
Incorporating community input and ensuring inclusivity in AI model design are essential steps to address existing gaps in GenAI applications and foster the development of equitable, context-sensitive technologies. Cultural and social contexts play a crucial role in this dynamic, underscoring the need for community involvement and multidisciplinary approaches to mitigate biases.  
The emerging field of Participatory AI \cite{b16} offers promising opportunities for both GenAI development and community advancement, fostering mutual empowerment and leading to the creation of more inclusive technologies. Participatory AI aligns closely with the principles of PRA, emphasizing the collaborative process of empowering communities and amplifying the voices of marginalized groups in decision-making. By actively involving communities in shaping the technologies that affect their lives, Participatory AI creates a reciprocal relationship where communities influence AI systems, and AI, in turn, addresses their specific needs. This is particularly important for gender research, as it opens pathways to redefine empowerment by integrating women’s insights and addressing context-specific inequities. Ultimately, aligning with the principles of Participatory AI facilitates mutual empowerment between AI systems and communities, paving the way for more inclusive and context-aware technologies that can be practically leveraged for solving society’s most pressing problems.

\section*{Acknowledgment}
The authors gratefully acknowledge Amrita Vishwa Vidyapeetham and its Chancellor, Sri Mata Amritanandamayi Devi, for the guidance, support, and inspiration provided throughout this research. Her vision of education for life and emphasis on compassion-driven research continues to guide our work in technology for sustainable development.

\section*{Conflict of Interest}
The authors declare no conflict of interest.

\end{document}